\documentclass[11pt]{article}
\usepackage[american]{babel}
\usepackage{fullpage}
\usepackage{amsmath}
\usepackage{amssymb}
\usepackage[title,titletoc]{appendix}
\usepackage{hyperref}
\usepackage{color}

\newcommand{\heading}[1]{\begin{center} {#1} \end{center}}

\newcommand{\spc}[1]{\hspace{#1cm}}
\newcommand{\sspc}{\spc{0.25}}
\newcommand{\lspc}{\spc{0.5}}

\newcommand{\longcomma}{\spc{0.3} , \spc{0.7}}
\newcommand{\psik}{\spc{0.3} ,}
\newcommand{\nekuda}{\spc{0.3} .}
\newcommand{\spsik}{\spc{0.1} ,}
\newcommand{\snekuda}{\spc{0.1} .}
\newcommand{\nn}{\nonumber}

\newcommand{\rr}{r_{23}}
\newcommand{\tot}{\theta_{12}}
\newcommand{\toth}{\theta_{13}}
\newcommand{\ttth}{\theta_{23}}

\newcommand{\e}{\varepsilon}
\newcommand{\Me}{M_E}
\newcommand{\Mnu}{M_\nu}

\newcommand{\B}{black}
\newcommand{\W}{white}

\newcommand{\msymm}[6]{{\tiny \left(\begin{array}{ccc}
\textcolor{#1}{\bullet} & \textcolor{#4}{\bullet} & \textcolor{#5}{\bullet} \\
\textcolor{#4}{\bullet} & \textcolor{#2}{\bullet} & \textcolor{#6}{\bullet} \\
\textcolor{#5}{\bullet} & \textcolor{#6}{\bullet} & \textcolor{#3}{\bullet} \end{array}\right)}}

\begin{document}

\heading{\LARGE \bf Abelian symmetries as the source of the small neutrino-related flavor parameters}
\vspace{0.3cm}

\renewcommand{\thefootnote}{\fnsymbol{footnote}}
\heading{\Large Shahar Amitai\footnote{shahar.amitai@weizmann.ac.il}}
\renewcommand{\thefootnote}{\arabic{footnote}}
\addtocounter{footnote}{-1}

\heading{\large \emph{Department of Particle Physics and Astrophysics, \\
Weizmann Institute of Science, Rehovot 76100, Israel}}
\vspace{0.5cm}

\begin{abstract}
There are four neutrino-related flavor parameters that have been measured: The three mixing angles $s_{23}$, $s_{12}$, $s_{13}$, and the ratio of mass-squared differences $\rr \equiv \Delta m_{21}^2 / |\Delta m_{31}^2|$. Of these, the first two are order one. On the other hand, $s_{13}$ and $\rr$ can be either order-one parameters that are accidentally somewhat small, or they are small for a reason, for example, they vanish in the limit of a symmetry that is broken by small parameters. We show that in the latter case, the Froggatt-Nielsen mechanism could explain the smallness of $s_{13}$ and $\rr$ only if some order-one coefficients are as small as the symmetry-breaking parameters. It is thus very unlikely that an Abelian symmetry is responsible for the smallness of $s_{13}$ and $\rr$.
\end{abstract}
\vspace{0.3cm}

\section{Introduction} \label{sec:int}
The flavor parameters in the quark sector -- quark masses and CKM mixing angles -- exhibit smallness and hierarchy. This structure is special enough that it is unlikely to be accidental. A leading candidate to explain these features is an approximate symmetry. The smallness is explained by the vanishing of all the small parameters in the symmetry limit. The hierarchy is explained by different powers of the small symmetry-breaking parameter(s) entering the various Yukawa couplings. The simplest successful attempt to explain the smallness and hierarchy in the quark flavor parameters is provided by the Froggatt-Nielsen mechanism \cite{Froggatt:1978} which employs a $U(1)$ symmetry. It predicts several relations among the flavor paremeters that are all consistent with the measured values \cite{Leurer:1992}.

The situation in the lepton sector is less clear. The recent measurements of $|U_{e3}|$ \cite{An:2012, Ahn:2012, Abe:2011a, Abe:2011b, Adamson:2011} have extended the list of measured lepton flavor parameters to eight: Three mixing angles, two neutrino mass-squared differences, and three charged lepton masses. Of these, there are four dimensionless parameters that seem to be small (See \cite{Fogli:2012, Tortola:2012, GonzalezGarcia:2012} for the neutrino parameters, and \cite{Beringer:1900} for the charged lepton masses):
\begin{align}
s_{13} \equiv \sin\toth &= 0.15 \spsik \label{eq:s13} \\
\rr \equiv \Delta m^2_{21} / |\Delta m^2_{31}| &= 0.031 \spsik \label{eq:r23} \\
m_\mu / m_\tau &= 0.059 \spsik \label{eq:rmt} \\
m_e / m_\mu &= 0.0048 \snekuda \label{eq:ret}
\end{align}
We quote here only the central values, because the experimental errors are small enough that they cannot change our analysis. The two other mixing angles, $s_{23} \equiv \sin\ttth$ and $s_{12} \equiv \sin\tot$, are not small. One may wonder why we do not list $m_\tau / v$ (with $v \simeq 246$ GeV the vacuum expectation value (VEV) of the Standard Model (SM) Higgs) as a small flavor parameter. The reason is that the smallness of $m_\tau / v$ may be the result of a small VEV $v_\ell$ of the scalar that couples to the charged leptons, $v_\ell \sim m_\tau$. Of course, if, instead, it is a result of a small tau-Yukawa, it should be added to the above list. Similarly, we do not consider each neutrino mass-squared difference as a small flavor parameter, because the overall scale of the neutrino masses is likely to be related to the seesaw mechanism rather than to flavor physics; We do not know whether the neutrino masses are small compared to the scale of $v^2 / \Lambda_{\rm seesaw}$ or not.

The issue of flavor model building depends crucially on how one interprets the data. In the context of an approximate symmetry, one has to decide which observables are truly small and therefore must vanish in the symmetry limit, and which ones are order-one parameters (which may accidentally assume somewhat small values), not suppressed by powers of symmetry-breaking parameters. Our own judgement is that the charged lepton mass ratios, $m_e / m_\mu$ (\ref{eq:ret}) and $m_\mu / m_\tau$ (\ref{eq:rmt}), are small, and that therefore $m_e$ and $m_\mu$ should vanish in the symmetry limit. On the other hand, the value of $s_{13}$ (\ref{eq:s13}) is large enough that both options -- accidental smallness and parametric suppression -- are possible.

As concerns $\rr$ (\ref{eq:r23}), things depend on whether neutrino masses have normal hierarchy, inverted hierarchy, or quasi-degeneracy. For normal hierarchy, the relevant flavor parameter is $m_2 / m_3 \approx \sqrt{\rr} \simeq 0.18$, comparable to $s_{13}$ and possibly just accidentally small. With inverted hierarchy, the relevant flavor parameter is $\Delta m_{21}^2 / m_2^2 \simeq \rr$, close to the charged lepton mass ratios and probably truly small. In the case of quasi-degeneracy, $\Delta m^2_{21} / m_2^2 \ll \rr$ and should definitely be parametrically suppressed. In all cases, models in which the smallness of $s_{13}$ is explained from symmetry reasons, but the smallness of $\rr$ is put by hand, make little sense to us.

Our aim in this work is to answer a simple question: Can an approximate Abelian symmetry account for the smallness of all four lepton flavor parameters of Eqs. (\ref{eq:s13})-(\ref{eq:ret})? If the answer is in the affirmative, then it is interesting to identify the relevant class of models and to find whether they lead to further predictions. If not, then we would be led to consider two possibilities: Either the explanation to smallness and hierarchy does not lie in the framework of approximate Abelian symmetries, or the somewhat small values of $s_{13}$ and $\sqrt{\rr}$ are just accidentally so.

\section{The framework} \label{sec:ang}
The leptonic mass terms take the following form:
\begin{align}
{\cal L}_{\rm mass}^{\rm leptons} = \nu_i (\Mnu)_{ij} \nu_j + {\overline{\ell_L}}_i (\Me)_{ij} {\ell_R}_j + {\rm h.c.} \psik
\end{align}
where $i$ and $j$ are flavor indices, $\Mnu$ is symmetric, and we take the neutrinos to be of Majorana type. The two mass matrices can be diagonalized as follows:
\begin{align}
U_\nu^T \Mnu U_\nu = D_\nu \longcomma U_\ell^\dagger \Me \Me^\dagger U_\ell = D_e^2 \psik
\end{align}
where $D_\nu = {\rm diag}(m_1, m_2, m_3)$ and $D_e = {\rm diag}(m_e, m_\mu, m_\tau)$. The leptonic mixing matrix is then given by
\begin{align}
U = U_\ell^\dagger U_\nu \nekuda
\end{align}

We assume that the entries of $\Mnu$ and $\Me$ are subject to selection rules that arise from an approximate continuous Abelian symmetry. We denote the charges of the $SU(2)$-doublet leptons under this symmetry by $H^L_i$, and those of the charged $SU(2)$-singlet leptons by $H^R_i$. As long as we are not interested in the structure of the scalar potential, we can always choose the charges of the relevant Higgs fields to be zero. In the symmetry limit,
\begin{align} \label{eq:condMnu}
(\Mnu)_{ij} = 0 \lspc {\rm if} \lspc H^L_i + H^L_j \neq 0
\end{align}
(and order one otherwise). Thus, in the symmetry limit, a diagonal entry would be different from zero only if the corresponding field carries charge zero, while an off-diagonal entry would be different from zero only if the sum of charges of the corresponding fields is zero. This simple observation implies that, in the symmetry limit, there are seventeen possible structures for $\Mnu$: one where all entries are allowed (all $H^L_i = 0$), one where all entries vanish, and five classes of other forms, each standing for three matrices that are related by flavor permutations. For each of the seventeen forms there is a large set of possible $\Me$ where, in the symmetry limit,
\begin{align} \label{eq:condMe}
(\Me)_{ij} = 0 \lspc {\rm if} \lspc H^L_i - H^R_j \neq 0
\end{align}
(and order one otherwise). Let us clarify that we use the term ``order one" to mean that these parameters are not suppressed by the small symmetry-breaking parameters, though their exact numerical value is not determined by the symmetry.

\section{Explaining $\rr \ll s_{13} \ll 1$} \label{sec:small}
One of the main challenges in lepton flavor model building, when assuming $s_{13} = 0$, is to find a symmetry that allows two mixing angles of order one, but makes the third mixing angle vanish. As a first stage, we examine all types of symmetries specified above, and check which ones produce a single vanishing mixing angle in the symmetry limit. In other words, we examine the 17 forms of $\Mnu$ and, for each of them, all the relevant forms of $\Me$. For each type of symmetry, we find $U_\nu$ and $U_\ell$ by diagonalizing the most general $\Mnu$ and $\Me \Me^\dagger$, respectively, and then obtain the mixing matrix $U$. Note that permuting rows or columns of $U$ corresponds to reordering the charged lepton or neutrino masses. Therefore, if we obtain a zero entry in $U$, it can always be moved to be $U_{e3}$. In addition, whenever we have a degenerate subspace (in terms of the neutrino or charged lepton masses) we have the freedom to rotate it by an arbitrary angle. In these cases, without loss of generality, we choose $U_\nu$ or $U_\ell$ that gives the largest number of vanishing mixing angles \footnote{Indeed, the degeneracy of the three neutrino masses within the SM ($\Mnu = 0$) is the reason for the absence of lepton mixing within this model.}.

Going through this procedure, we obtain a single vanishing mixing angle for two types of symmetries:
\begin{itemize}
\item {\bf Type I:} Symmetries with $H_e^L = +1$, $H_\mu^L = H_\tau^L = -1$, and at least one of the $H_i^R$ charges equal to $-1$ \footnote{This is up to a flavor permutation: $(H_e^L, H_\mu^L, H_\tau^L) = (-1, +1, -1)$ or $(-1, -1, +1)$ also belong to this type.}. These symmetries give $\Me$ of various forms, but the same $\Me \Me^\dagger$:
\begin{align} \label{eq:forms}
\Mnu = \msymm{\W}{\W}{\W}{\B}{\B}{\W} \longcomma \Me \Me^\dagger = \msymm{\B}{\B}{\B}{\W}{\W}{\B} \psik
\end{align}
where the empty entries vanish in the symmetry limit, and the others are of order one. For symmetries with no $H_i^R = +1$, we have also $(\Me \Me^\dagger)_{11} = 0$. All these symmetries lead to $\toth = 0$ and a maximal $\tot$. It is easy to reach this conclusion starting from Eq. (\ref{eq:forms}): An appropriate $2-3$ rotation of $\Mnu$ will make $(\Mnu)_{13/31}$ vanish, and then a maximal $1-2$ rotation is needed to make $\Mnu$ diagonal. Obviously, $\Me$ only needs a $2-3$ rotation to make it diagonal, so overall we get $\toth = 0$ and $\tot = \pi / 4$ \footnote{We are using the common convention $U = R_{23} (\ttth) R_{13} (\toth) R_{12} (\tot)$.}.
\item {\bf Type II:} Symmetries with vanishing left-handed charges ($H_e^L = H_\mu^L = H_\tau^L = 0$), and with only two non-vanishing right-handed charges ($H_e^R = 0$ or $H_\mu^R = 0$ or $H_\tau^R = 0$). These symmetries give $\Mnu$ completely general and $\Me$ with only one non-vanishing column. Since we have a $2 \times 2$ degenerate subspace for the charged leptons, we can make one of the mixing angles vanish. The two other mixing angles are unknown.
\end{itemize}

We next require that the Abelian symmetry imposes $\rr = 0$ in the symmetry limit. Notice that the value of $\rr$ only depends on $\Mnu$, namely on the left-handed charges. For Type I symmetries we get $m_1 = - m_2$ and $m_3 = 0$, which indeed gives $\rr = 0$. For Type II symmetries, the form of $\Mnu$ is completely general, so $\rr$ is expected to be of order one, and therefore these symmetries do not explain its smallness.

We now move away from the symmetry limit, and check whether the measured values of the small parameters $\rr$ and $s_{13}$ can be accounted for by the small symmetry-breaking parameters. Take, for example, the well-known Type I (vector-like) symmetry:
\begin{align} \label{eq:knownsymm}
H = L_e - L_\mu - L_\tau \psik
\end{align}
where $L_i$ is the corresponding lepton flavor charge:
\begin{align}
(H_e^L, H_\mu^L, H_\tau^L) = (H_e^R, H_\mu^R, H_\tau^R) = (+1, -1, -1) \nekuda
\end{align}
If this symmetry is approximate at low energy, we obtain
\begin{align} \label{eq:FOforms}
\Mnu = m_\nu \left( \begin{array}{ccc} \e' & 1  & 1  \\
                                       1   & \e & \e \\
                                       1   & \e & \e \end{array} \right) \longcomma
\Me  = m_e   \left( \begin{array}{ccc} 1   & \e & \e \\
                                       \e' & 1  & 1  \\
                                       \e' & 1  & 1  \end{array} \right) \nekuda
\end{align}
For each entry in either mass matrix we only specify the parametric suppression, and omit arbitrary order-one coefficients. $\e$ and $\e'$ are spurions, namely small symmetry-breaking parameters of $H$-charges $+2$ and $-2$, respectively. Using these mass matrices, we get first-order corrections to the predictions derived in the symmetry limit:
\begin{align} \label{eq:predictions}
\toth \sim \e + \e' \longcomma \frac{\pi}{4} - \tot \sim \e + \e' \longcomma \rr \sim 4 (\e + \e') \nekuda
\end{align}
Note that the neutrino spectrum is of the inverted hierarchy type and, therefore, as explained in the introduction, the relevant related small parameter is indeed $\rr$ (and not $\sqrt{\rr}$). The current experimental ranges of these parameters imply
\begin{align}
\toth \sim 0.15 \longcomma \frac{\pi}{4} - \tot \sim 0.20 \longcomma \rr \sim 0.03 \nekuda
\end{align}
We see a mismatch between the values of the two mixing angles and the value of $\rr$. While $\toth$ and $\tot$ require symmetry-breaking parameters of the order $0.1-0.2$, the smallness of $\rr$ requires the same symmetry-breaking parameters to be of order 0.01 (Notice the factor of 4 in Eq. (\ref{eq:predictions})). If we want all three parameters to assume their experimental values, we need accidental numbers of order one to be as small as the small symmetry-breaking parameters (or as large as the inverse of the small breaking parameters). Under such circumstances, it is fair to conclude that the symmetry (\ref{eq:knownsymm}) fails to give a natural explanation for the small parameters $\toth$ and $\rr$ \footnote{For previous analyses of the symmetry $L_e - L_\mu - L_\tau$, see \cite{Nir:2000, Barbieri:1998}}.

Our statement that the pattern of the first-order corrections to the symmetry limit is inconsistent with phenomenology was based on the specific example of the symmetry (\ref{eq:knownsymm}). Can other Type I symmetries overcome the problem? The form of $\Mnu$ (\ref{eq:FOforms}) is common to all these symmetries, so the parametric suppression of $\rr$ is the same. The various Type I symmetries may differ, however, in the first-order form of $\Me \Me^\dagger$ (The zeroth-order form (\ref{eq:forms}) is the same). In particular, first-order corrections to the mixing angles coming from $U_\ell$ can be different, or even absent. But since $\Mnu$ leads to first-order corrections of its own, the problem appears for all these symmetries. We learn that all Type I symmetries fail to explain the fact that $\rr \sim s_{13}^2$.

Our assumption that all vanishing mixing angles in the symmetry limit are small when the symmetry is broken does not always hold. In the case where all entries of $\Mnu$ vanish ($H^L_i \neq 0$ and $H^L_i + H^L_j \neq 0$ for all $i$, $j$), order-one mixing angles can be obtained from the leading order of the symmetry-breaking parameter. But if all left-handed charges are of the same sign, the leading-order structure is one of the 17 we have analyzed. The form of $\Me$ comes from the corresponding set of possible $\Me$, and therefore these symmetries are included in our analysis. In the case where all entries of $\Me$ vanish ($H^L_i - H^R_j \neq 0$ for all $i$, $j$), we give a similar argument: Given a particular structure of $\Mnu$, the set of entries in $\Me$ that are the same order of the symmetry-breaking parameter always form a structure that has been analyzed. Therefore, assuming a single spurion, we get no new leading-order structures for $\Me$.

The additional requirement that the symmetry explains the smallness of at least two of the charged lepton masses, Eqs. (\ref{eq:rmt}) and (\ref{eq:ret}), further constrains the list of possible symmetries. In particular, it dictates charges of the $SU(2)$-singlet charged leptons. For example, for the Type I symmetry with $(H_e^L, H_\mu^L, H_\tau^L) = (+1, -1, -1)$ to generate two massless charged leptons (in the symmetry limit), one (and only one) of the right-handed charges $H_i^R$ must equal $-1$, while the other two charges must not equal $+1$. These extra constraints, however, have no effect on our conclusion of the failure of these symmetries to account for $\rr \ll s_{13} \ll 1$.

So far, we studied $U(1)$ symmetries. One may wonder whether more elaborate continuous symmetries or, more specifically, $[U(1)]^n$ symmetries, can lead to different conclusions. When we combine two symmetries, the allowed entries of the mass matrices (in the symmetry limit) are the intersection of the two sets of allowed entries. Since a larger number of zero entries can only reduce the number of non-vanishing mixing angles, there is no need to consider any $U(1)$ factor that leads to more than a single vanishing mixing angle on its own. We should only consider $U(1)$ factors that lead to one vanishing mixing angle or none. We have listed above two types of $U(1)$ symmetries that lead to a single vanishing mixing angle (Type I and Type II). When considering $U(1)$ factors that lead to no vanishing mixing angle, we are led to define yet another class of symmetries:
\begin{itemize}
\item {\bf Type III:} This class includes the three Abelian symmetries $L^R_i$, defined by having the corresponding $H^R_i \neq 0$ and all other charges zero. By themselves, Type III symmetries give $\Mnu$ completely general and $\Me$ with one vanishing column.
\end{itemize}
Any combination of Type II and/or Type III symmetries leaves $\Mnu$ completely general, and therefore fails to account for $\rr \ll 1$. To account for $\rr \ll 1$, Type I symmetries must be involved. Adding Type II or Type III symmetry to Type I symmetry sets columns of $\Me$ to zero. In some cases, this action adds another vanishing mixing angle (in the symmetry limit), and therefore these particular combinations do not work. In other cases, this action changes the symmetry-limit form of $\Me$ to another Type I form. Yet, these combinations do not change the form of $\Mnu$, so $\rr$ gets first-order corrections. Therefore, these combinations still fail to account for $\rr \ll \toth \ll 1$.

We are left with the option of combining two Type I symmetries (Having more than two $U(1)$'s does not change the argument below). There are two possible scenarios here. The first scenario involves two $U(1)$'s with different left-handed charges (so each of these $U(1)$ factors by itself gives a different form for $\Mnu$). In the symmetry limit, $\Mnu$ has only two allowed entries, while $\Me \Me^\dagger$ is diagonal. Consequently, only one non-vanishing mixing angle is generated. Thus, this scenario is excluded. The second scenario involves two $U(1)$'s with the same left-handed charges (same form for $\Mnu$). In this case, $\Mnu$ only gives second-order corrections to the mixing angles and to $\rr$ and $\Me$ only gives second-order corrections to the mixing angles. Consequently, the parametric suppression of $\toth$ and $\rr$ is still of the same order and the problem remains.

The bottom line is that no low-energy effective continuous Abelian symmetry can give a natural explanation for the small parameters $\toth$ and $\rr$.

\section{Discrete Abelian symmetries} \label{sec:discrete}
In the case of a discrete Abelian symmetry $Z_n$, we have, without loss of generality, $H^{L,R}_i \in \mathbb{Z}_n$. In the symmetry limit,
\begin{align}
(\Mnu)_{ij} = 0 \lspc {\rm if} \lspc H^L_i + H^L_j \neq 0 \sspc ({\rm mod} \sspc n)
\end{align}
(and order one otherwise). Since a discrete transformation is a particular case of a continuous one, all matrix structures specified in section \ref{sec:int} for a continuous symmetry can be a result of a discrete symmetry too. A discrete symmetry, though, can produce two additional structures for $\Mnu$ (times three flavor permutations) coming from the fact that a diagonal entry can also be different from zero if the corresponding field carries charge $n / 2$. For each of these two additional forms there is a large set of possible $\Me$ that, in the symmetry limit, satisfy eq. (\ref{eq:condMe}) \footnote{Choosing $H^{L,R}_i \in \mathbb{Z}_n$ like we did, the only option for the subtraction of two charges to be zero (mod $n$) is zero.}. Including these extra structures in our analysis, we get no additional types of symmetries that lead to one vanishing mixing angle, or none, in the symmetry limit. We conclude that no low-energy effective Abelian symmetry (continuous, discrete, or a combination of the two) can give a natural explanation for the small parameters $\toth$ and $\rr$.

\section{Seesaw mechanism} \label{sec:see}
So far we have assumed the selction rules that come from an approximate Abelian symmetry apply directly to the low-energy effective Lagrangian and, in particular, to the light neutrino mass matrix. In this section we assume that the Majorana mass terms come from a seesaw mechanism and consider the case that the symmetry acts on the full high energy theory, which includes the $SU(2)$-singlet neutrinos. In such a scenario, the effective neutrino mass matrix $\Mnu$ is not necessarily the most general matrix allowed by the approximate symmetry \cite{Barbieri:1998}. $\Mnu$ would be the most general matrix allowed by the approximate symmetry when all singlet neutrino charges have the same sign, and there is no more than one spurion breaking the symmetry \cite{Grossman:1995, Altarelli:2004}. In terms of the form of $\Mnu$, these cases are equivalent to a low-energy effective theory, and therefore all the previous results hold.

\section{Neutrino flavor anarchy} \label{sec:anarchy}
Our statement that the Froggatt-Nielsen mechanism cannot account for the measured values of $s_{13}$ and $\rr$ depends crucially on our interpretation of the data as implying that these parameters are parametrically suppressed. The other possible interpretation, that these (or, more precisely, $s_{13}$ and $\sqrt{\rr}$) are order-one parameters that are accidentally suppressed by order-one coefficients, is consistent with a $U(1)$ symmetry that suppresses the charged lepton masses, but leaves all neutrino-related parameters unsuppressed. Such a scenario, known as ``Anarchy" \cite{Hall:1999, Haba:2000}, requires simply $H_e^L = H_\mu^L = H_\tau^L$. The hierarchy in the charged lepton masses is explained by, for example, $H_e^R > H_\mu^R > H_\tau^R \geq 0$.

\section{Conclusions} \label{sec:con}
We examined all possible Abelian symmetries that dictate selection rules to the lepton mass matrices, with the aim of explaining the four small parameters of the lepton sector in a natural way. We focussed on symmetries that predict $\toth = 0$ and $\rr \equiv \Delta m^2_{21} / |\Delta m^2_{31}| = 0$ in the symmetry limit. We found that this symmetries lead to inverted hierarchy and, in the symmetry limit, to maximal $1-2$ mixing. We further found that all these models predict $\rr$, $\toth$ and $\pi / 4 - \tot$ of the same parametric suppression, while in reality $\rr$ is much smaller than $\toth$ and than $\pi / 4 - \tot$. This situation can only be fixed at the price of introducing order-one parameters that are accidentally as small as the symmetry-breaking parameters. Under such circumstances, it is hard to see the motivation to think that these models apply in Nature.

During recent years, lepton flavor models were evaluated by numerically estimating their ``success rate" to produce small enough values for $s_{13}$ and $\rr$. In particular, models with neutrino flavor anarchy were evaluated \cite{Hall:1999, deGouvea:2012}, and were also compared to models that enforce neutrino hierarchy by using the Froggatt-Nielsen mechanism \cite{Altarelli:2002, Altarelli:2012}. The success rate of producing both $s_{13}$ and $\rr$ smaller than their current measured values is of order 1\% at the most. Our results explain the difficulty in achieving a higher success rate.

Our conclusion leaves open two different possibilities within the assumption that the structure in the flavor parameters is the consequence of an approximate symmetry:
\begin{itemize}
\item Approximate $U(1)$ symmetries (the Froggatt-Nielsen mechanism) are at work, but the interpretation of $\sqrt{\rr}$ and $|U_{e3}|$ as small parameters is incorrect. In other words, there is neutrino flavor anarchy.
\item The symmetry that dictates the structure of the flavor parameters is non-Abelien.
\end{itemize}

\section*{Acknowledgments}

I thank Yossi Nir for useful discussions and assistance with the draft. I thank Efrat Gerchkovitz and Tali Vaknin for many useful discussions. The research of SA is supported by the Israel Science Foundation.

\begin{appendices}

\section{$\Mnu$ with $|U_{e3}| = 0$}

For the purposes of this appendix, we work in the flavor basis (where $\Me$ is diagonal). In this basis, the mixing matrix is made of the three eigenvectors $\vec{U}$ of $\Mnu$:
\begin{align}\label{eq:ev}
\Mnu \vec{U} = \lambda \vec{U} \nekuda
\end{align}
Having $\toth = 0$ (or $|U_{e3}| = 0$) means that one of the eigenvectors is of the form
\begin{align}
\vec{U} = \left(\begin{array}{c} 0 \\ 1 \\ a \end{array}\right) \nekuda
\end{align}
Eq. (\ref{eq:ev}) provides three relations among the entries of $\Mnu$. Since we introduced two new variables - $a$ and $\lambda$, we should expect to get one condition on the entries that is independent of $a$ and $\lambda$. Indeed, solving the three equations we get the condition:
\begin{align} \label{eq:rel}
m_{33} = m_{22} + m_{32} \left( \frac{m_{31}}{m_{21}} - \frac{m_{21}}{m_{31}} \right) \nekuda
\end{align}

We have found a way to identify mass matrices that lead to $\toth = 0$. Eq. (\ref{eq:rel}) is satisfied by all these matrices, and only by them \footnote{Notice that eq. (\ref{eq:rel}) cannot be satisfied when $m_{21} = 0$ or $m_{31} = 0$. In this case, the solution for the three equations is $m_{21} = m_{31} = 0$. This means that at least two mixing angles vanish, so it is not an interesting scenario.}. It is indeed satisfied by the mass matrix in a popular $A4$-based model \cite{Altarelli:2005} and also by the mass matrix resulting from the $L_e - L_\mu - L_\tau$ symmetry discussed above:
\begin{align}
\Mnu^{(A4)} \sim \left( \begin{array}{ccc} A + 2B &   -B &   -B \\
                                               -B &   2B & A -B \\
                                               -B & A -B &   2B \end{array}\right) \longcomma
\Mnu^{(L_e-L_\mu-L_\tau)} \sim \left( \begin{array}{ccc} 0 & A & B \\
                                                         A & 0 & 0 \\
                                                         B & 0 & 0 \end{array}\right) \nekuda
\end{align}

We can express the constraint of Eq. (\ref{eq:rel}) in a more convenient manner. For example, the matrices
\begin{align}
\Mnu^{(1)} = \left( \begin{array}{ccc} A                & B              & B \sin{\theta}       \\
                                       B                & C              & D \sin{\theta}       \\
                                       B \sin{\theta}   & D \sin{\theta} & C - D \cos{\theta}^2 \end{array}\right) \longcomma
\Mnu^{(2)} = \left( \begin{array}{ccc} A                & B              & B / \sin{\theta}     \\
                                       B                & C              & D \sin{\theta}       \\
                                       B / \sin{\theta} & D \sin{\theta} & C + D \cos{\theta}^2 \end{array}\right) \nn
\end{align}
are the most general matrices that satisfy eq. (\ref{eq:rel}). Indeed, they have five free parameters instead of six. We can use one of them, depending on which entry is bigger: $m_{21}$ or $m_{31}$ \footnote{In terms of the three mixing angles, we get, for $\Mnu^{(1)}$, $\sin{\theta} = \tan{\ttth}$, or, for $\Mnu^{(2)}$, $\sin{\theta} = \cot{\ttth}$. Indeed, when $\sin{\theta} = 1$ we get the most general mass matrix that corresponds to $\toth = 0$ and a maximal $\ttth$.}.

\end{appendices}

\bibliography{AbelianBib}
\bibliographystyle{utphys}

\end{document}